\RequirePackage{fix-cm}
\documentclass[smallextended]{svjour3}       
\smartqed  
\usepackage{graphicx}
\usepackage{amsmath}
\usepackage{indentfirst}
\usepackage{amsfonts,amssymb}
\usepackage{mathrsfs}
\usepackage{lineno,hyperref}
\usepackage{tikz}
\usetikzlibrary{shapes,arrows}
\newtheorem{Definition}{Definition}[section]
\newtheorem{Theorem}{Theorem}[section]

\newtheorem{Proposition}{Proposition}[section]
\newtheorem{Example}{Example}[section]
\numberwithin{equation}{section}
\newtheorem{Remark}{Remark}[section]
\begin{document}

\title{Set-valued risk statistics with the time value of money
}


\author{Fei Sun*        \and Xiaozhi Fan  \and Weitao Liu 
}


\institute{*Corresponding author\\
	F. Sun \at
              	School of Mathematics and Computational Science, Wuyi University, Jiangmen 529020, China \\
              \email{fsun.sci@outlook.com}           
          \and 
          XZ Fan \at
          Changjiang Institute of Survey, Planning, Design and Research, Wuhan 430010, China \\
          \email{fanxiaozhi@cjwsjy.com.cn}          
               \and
             WT Liu \at 
            Guangzhou Institute of Urban Strategy, Guangzhou Academy of Social Sciences, Guangzhou 510410, China \\
            \email{cyllwt@aliyun.com}
}

\date{
}

\maketitle

\begin{abstract}
The time value of money is a critical factor not only in risk analysis, but also in insurance and financial applications. In this paper, we consider a special class of set-valued risk statistics by introducing the time value of money.
In fact, the risk statistics established by this method is closer to financial reality than traditional ones.
Moreover, this new risk statistic can be uesd for the quantification of portfolio risk. By further developing the properties related to these risk statistics, we are able to derive representation results for such risk.
\keywords{risk statistics \and  set-valued \and portfolio \and time value}

\end{abstract}

\section{Introduction}
\label{sec:1}
Research on risk is a popular topic in both quantitative and theoretical research, and risk models have attracted considerable attention.
The quantitative calculation on risk involves
two problems: choosing an appropriate risk model and allocating risk to individual institutions. This has led to further research on risk statistics.\\

In traditional research of risk statistic, the property of translative invariance denies the time value of money. 
However, as pointed out by EL Karouii and Ravanelli (2009), the translative invariance axiom may fail once there is any form of uncertainty about interest rates because the money has a time value. For example, when $m$ dollars are added to a future position $X$, the capital requirement at time $t=0$ is reduced by less than $m$ dollars because the value of the money may grow with time. 
Therefore, it is more appropriate to study risk statistics from the perspective of the time value of money.

 Evaluating risk of a portfolio consisting of several financial positions. And the set-valued risk measures are more appropriate than scalar risk measures, especially in the case where several different kinds of currencies are involved when one is determining capital requirements for the portfolio. In fact, Hamel and Heyde (2010) pointed out that the basic question for quantifying portfolio risk is how to evaluate the risk of a multivariate random outcome in terms of more than one reference instrument, for example if the regulator accepts deposits in more than one currency. This is of particular importance if transaction costs have to be paid for each transaction between assets including the reference instruments.
Other studies of set-valued risk measures include those of Hamel (2009), Hamel et al. (2011), Hamel et al. (2013), Labuschagne and Offwood-Le Roux (2014), Farkas et al. (2015), Molchanov and Cascos (2016), Ararat et al. (2017), Deng and Sun (2020), Sun and Dong (2021) and the references therein. A natural set-valued risk statistic can be considered as an empirical (or a data-based) version of a set-valued risk measure. 

From the statistical point of view, the behaviour of a random variable can be
characterized by its observations, the samples of the random variable. Heyde et al. (2007) and Kou et al. (2013) first
introduced the class of natural risk statistics and the corresponding
representation results are also derived. 
Later, Tian and Suo (2012) obtained representation results for convex risk statistics,
and the corresponding results for quasiconvex risk statistics were obtained by Tian and Jiang (2015).
However, all of these risk statistics are designed to quantify risk of  a single financial position (i.e. a random variable) by its samples. A natural question is determining how to quantify risk of a portfolio by its samples, especially in the situation where different kinds of currencies are possibly involved in the portfolio.

The main focus of this paper is a new class of set-valued risk statistics with the time value of money, named cash sub-additive risk statistics with an axiomatic approach. By further developing the properties related to cash sub-additive risk statistics, we are able to derive representation results for such risk. This new class of risk statistics can be considered as an extension of those introduced by Sun and Hu (2019) from the empirical (or a data-based) version.

	\tikzstyle{startstop} = [rectangle,rounded corners, minimum width=3cm,minimum height=1cm,text centered, draw=black]
	\tikzstyle{io} = [trapezium, trapezium left angle = 70,trapezium right angle=110,minimum width=3cm,minimum height=1cm,text centered,draw=black]
	\tikzstyle{process} = [rectangle,minimum width=3cm,minimum height=1cm,text centered,text width =3cm,draw=black]
	\tikzstyle{decision} = [diamond,minimum width=3cm,minimum height=1cm,text centered,draw=black]
	\tikzstyle{arrow} = [thick,->,>=stealth]
	
	\begin{tikzpicture}[node distance=2cm]
	\node (start) [startstop] {Time value of money};
	\node (decision1) [process,below of=start] {Cash sub-additivity};
	\node (process2a) [process,below of=decision1] {Relation between two risk statistics};
	\node (process2b) [process,right of=decision1,xshift=2cm] {Definition};
	\node (out1) [process,below of=process2a] {Representation result of the model};	
	\node (exam) [decision,below of=process2b]{Example};
	\node (stop) [startstop,below of=out1] {Data-based version};

	\draw [arrow] (start) -- (decision1);
	\draw [arrow] (decision1) -- node[anchor=east] {} (process2a);
	\draw [arrow] (decision1) -- node[anchor=south] {} (process2b);
	\draw [arrow] (process2b) |- (start);
	\draw [arrow] (process2b) -- (exam);
	\draw [arrow] (process2a) -- (out1);
	\draw [arrow] (out1) -- (stop);
	\end{tikzpicture}

The remainder of this paper is organized as follows. In Sect.~\ref{sec:2},
we briefly introduce some preliminaries. In Sect.~\ref{sec:3}, we state the representation result of cash sub-additive risk statistics.
In Sect.~\ref{sec:4}, we investigate the alternative data-based versions of cash sub-additive risk measures.
Finally, Sect.~\ref{sec:5} discusses the main proof in this paper.

\section{Preliminaries}
\label{sec:2}
In this section, we briefly introduce some preliminaries that are used throughout this paper. Let $d \geq 1$ be a fixed positive integer. The space $\mathbb{R}^{d\times n}$ represents the set of financial risk positions. an element $z$ of $\mathbb{R}^d$ is denoted by $ z:=(z_1, \cdots, z_d).$
An element $X$ of $\mathbb{R}^{d \times n}$ is denoted by
$
X:=(X_1, \cdots, X_d):=(x^{1, 1}_1, \cdots, x^{1, 1}_{n_1},$
$ \cdots, x^{1, l}_1, \cdots,  x^{1, l}_{n_l},
\cdots,  x^{d, 1}_1, \cdots x^{d, 1}_{n_1}, \cdots, x^{d, l}_1, \cdots,  x^{d, l}_{n_l}).
$
The $d\times n$ dimensional financial positions in $\mathbb{R}^{d\times n}$ have a strong realistic interpretation. This is indeed the case if we
consider realistic situations where investors have access to different markets and form multi-asset portfolios in the presence of frictions such as transaction costs, liquidity problems, irreversible transfers, etc. With positive values of $X\in \mathbb{R}^{d\times n}$ we denote the gains while the negative denote the losses. The behavior of the $d$-dimensional  random vector $ D =(X_1, \cdots, X_d)$ in different scenarios is represented by
different sets of data observed or generated in those scenarios because specifying accurate models for $D$ is usually very difficult. Here, we suppose that there always exist $l$ scenarios. Let $n_j$ be the sample size of $D$ in the $j^{th}$  scenario, $j=1, \cdots, l.$  Let $n:= n_1 +\cdots+n_l$. More precisely, suppose that the behavior of $D$
is represented by a collection of data $X=(X_1, \cdots, X_d) \in \mathbb{R}^n \times \cdots \times \mathbb{R}^n$, where
$X_i=(X^{i, 1}, \cdots, X^{i, l}) \in \mathbb{R}^n$,
$X^{i, j} =(x^{i, j}_1, \cdots, x^{i, j}_{n_j}) \in \mathbb{R}^{n_j}$ is the data subset that corresponds to the $j^{th}$ scenario with respect to $X_i$.
For each $j=1, \cdots, l$, $h=1, \cdots, n_j$, $X^j_h:=\left(x^{1, j}_h, x^{2, j}_h, \cdots, x^{d, j}_h\right)$  is the data subset that corresponds to the $h^{th}$ observation of $D$ in the $j^{th}$ scenario, and can be based on historical observations, hypothetical samples simulated according to a model, or a mixture of observations and simulated samples.\\

 Let $K$ be a closed convex polyhedral cone of $\mathbb{R}^{d}$ where $K\supseteq \mathbb{R}^{d}_{+}$ where $\mathbb{R}^{d}_{+}:=\{(x_{1},\ldots,x_{d})\in \mathbb{R}^{d}; x_{i}>0, 1\leq i\leq d\}$ and $K\cap \mathbb{R}^{d}_{-} = \emptyset$ where $\mathbb{R}^{d}_{-}:=\{(x_{1},\ldots,x_{d})\in \mathbb{R}^{d}; x_{i}\leq 0, 1\leq i\leq d\}$.  Let $K^{+}$ be the positive dual cone of $K$, that is $K^{+}:=\{u\in \mathbb{R}^{d}:u^{tr} v\geq0 \textrm{ for any } v\in K\}$, where $u^{tr}$ means the transpose of $u$. For any $X = (X_{1}, \ldots, X_{d}),  Y = (Y_{1}, \ldots, Y_{d})\in \mathbb{R}^{d\times n}$, $X + Y$ stands for $(X_{1}+Y_{1}, \ldots, X_{d}+Y_{d})$ and $aX$ stands for $(aX_{1}, \ldots, aX_{d})$ for $a\in \mathbb{R}$.
For any $z:=(z_1, \cdots, z_d) \in \mathbb{R}^d$, denote
$K1_n:=\{(z_{1} 1_n, z_{2} 1_n, \cdots, z_{d} 1_n): z \in K\}$ and  $z1_n:=\{(z, z, \cdots, z): z \in \mathbb{R}\}\in \mathbb{R}^n$ where $1_n := (1, \cdots, 1) \in \mathbb{R}^n$.
By $(K1_n)^+$ we denote the positive dual cone of $K1_n$ in $\mathbb{R}^{d \times n}$, i.e.
$(K1_n)^+:=\{w \in \mathbb{R}^{d \times n}: w z^{tr} \geq 0
\text{ for any } z \in K\}$. The partial order respect to $K$ is defined as $a\leq_{K} b$. Therefore, $b-a\in K$ where $a,b\in \mathbb{R}^{d}$ and $X\leq_{K1_n} Y$ means $Y-X\in K1_n$ where $X,Y\in \mathbb{R}^{d\times n}$.
Let $M:=\mathbb{R}^{m}\times \{0\}^{d-m}$ be the linear subspace of $\mathbb{R}^{d}$ for $1\leq m\leq d$. The introduction of $M$ was considered by Hamel (2009). Denote $M_{+}:=M\cap \mathbb{R}^{d}_{+}$ where $\mathbb{R}^{d}_{+}:=\{(x_{1},\ldots,x_{d})\in \mathbb{R}^{d}; x^{i}\geq0, 1\leq i\leq d\}$. Therefore, a regulator can only accept security deposits in the first $m$ reference instruments. Denote $K_{M}:=K\cap M$ by the closed convex polyhedral cone in $M$, $intK_{M}$ the interior of $K_{M}$ in $M$. We denote $\mathbb{T}_{M}:=\{A\subset M:A=clco(A+K_{M})\}$ and $\mathbb{T}_{M^{+}}:=\{A\subset K_{M}:A=clco(A+K_{M})\}$, where the $clco(A)$ represents the closed convex hull of $A$.\\

By Chen and Hu (2017), a set-valued convex risk statistic is any map 
that can be considered as an empirical (or a data-based) version of a set-valued risk measure. The  axioms related to this set-valued convex risk statistics $\rho: \mathbb{R}^{d\times n}\rightarrow \mathbb{T}_{M}$ are organized as follows,
\begin{description}
	\item[A0] Normalized: $K_{M}\subseteq \rho(0)$ and $\rho(0)\cap -intK_{M}=\phi$;
	\item[A1] Monotonicity: for any $X$,$Y\in \mathbb{R}^{d\times n}$, $X-Y\in K1_{n}$ implies that $\rho(X)\supseteq \rho(Y)$;
	\item[A2] Translative invariance: for any $X\in \mathbb{R}^{d\times n}$ and $z\in M$, $\rho(X-z1_{n})=\rho(X)+z$;
	\item[A3] Convexity: for any $X,Y\in \mathbb{R}^{d\times n}$ and $\lambda\in[0,1]$, $\rho(\lambda(X)+(1-\lambda)Y)\supseteq \lambda\rho(X)+(1-\lambda)\rho(Y)$.
\end{description}

  Let $D:=(D_1, \cdots, D_d)\in \mathbb{R}^{d\times n}$ be the stochastic discount factor for certain currency $X\in \mathbb{R}^{d\times n}$, i.e. each element of $D$ belones to $[0,1]$. A function $\rho: \mathbb{R}^{d\times n}\rightarrow \mathbb{T}_{M}$ is called proper if $\textrm{dom}\rho:=\{X\in \mathbb{R}^{d\times n}:\rho(X)\neq \emptyset \}\neq \emptyset $ and $\rho(X)\neq M$ for all $X\in \textrm{dom}\rho$. $\rho$ is said to be closed if  $\textrm{graph}\rho$ is a closed set with respect to the product topology in $\mathbb{R}^{d\times n}\times M$.\\

In fact, for the common convex risk statistics the translative invariance axiom has been largely accepted by academics and practitioners. However, as it pointed out by EL Karouii and Ravanelli (2009), while regulators and financial institutions determine and collect today the reserve amounts to cover future risky positions, the translative invariance axiom requires that risky positions and reserve amounts are expressed in the same num\'{e}raire. Unfortunately, when the interest rates are stochastic this procedure is unreasonable. Implicitly, since the time value of money, the cash sub-additivity turn out to be more suitable than the translative invariance axiom. In this paper, we will derive a new class of set-valued risk statistics with the time value of money.

\section{Empirical versions of set-valued cash sub-additive risk measures}
\label{sec:3}

Since the time value of money is a critical factor in risk analysis, we will consider a special class of set-valued risk statistics, named cash sub-additive risk statistics. Note that these set-valued risk statistics are the empirical versions of set-valued cash sub-additive risk measures introduced by Sun and Hu (2019).\\

In this section, we state the representation results of cash sub-additive risk statistics. However, our viewpoint is not the same as Chen and Hu (2017). Instead, we start from the relation between set-valued convex risk statistics and set-valued cash sub-additive risk statistics.\\ 

We begin with the axioms related to cash sub-additive risk statistics.

\begin{Definition}
		A set-valued cash sub-additive risk statistic is a set-valued function $R:\mathbb{R}^{d\times n}$ $\rightarrow$ $\mathbb{T}_{M}$  that satisfies $\mathbf{A0},\mathbf{A1},\mathbf{A3}$ and the following property.
		\begin{description}
			\item[A4] Cash sub-additivity: for any $X\in \mathbb{R}^{d\times n}$ and $z\in K_{M}$,
			\begin{displaymath}
			R(X+z1_{n})\subseteq R(X)-z  \qquad\textrm{or}\qquad  R(X-z1_{n})\supseteq R(X)+z.
			\end{displaymath}
		\end{description}
\end{Definition}

In fact, the cash sub-additivity of set-valued cash sub-additive risk statistic is derived from the idea of the time value of money. A basic reason is that while regulators and financial institutions determine and collect today the reserve amounts to cover future risky positions, the cash additivity requires that risky positions and reserve amounts are expressed in the same num\'{e}raire. Implicitly, when $m$ dollars are added to a future position $X$, the capital requirement at time $t=0$ is reduced by less than $m$ dollars because the value of the money may grow with time. In this context, we define a set-valued risk statistic in the discounted position.

\begin{Definition}
		Let $D\in \mathbb{R}^{d\times n}$ be the discount factor. A set-valued discount risk statistic, say $\overline{R}$, is a set-valued convex risk statistic which is defined in the discounted factor $DX$ where $X\in \mathbb{R}^{d\times n}$.
\end{Definition}

By set-valued discount risk statistic $\overline{R}$, we can define a set-valued convex function by $R(X):=\overline{R}(DX)$. As $\overline{R}$ is a set-valued convex risk statistic, for any $z\in K_{M}$, we have
\[
R(X+z1_{n})=\overline{R}(DX+Dz1_{n})\subseteq\overline{R}(DX+z1_{n})=\overline{R}(DX)-z=R(X)-z.
\]
This property of $R$ is due to the time value of money. That is to say that $R$ is expressed in terms of the current num\'{e}raire but defined in the future financial positions with the future num\'{e}raire. \\

\indent We now introduce a special case of set-valued cash sub-additive risk statistics that is said to be set-valued convex loss-based risk statistics, see Sun et al. (2018). Note that the scalar case of convex loss-based risk measures was studied by Chen et al.(2018) and Cont et al. (2013). The following definition is derived from Sun et al. (2018).\\

\begin{Definition}\label{D2}
		A set-valued loss-based risk statistic is a mapping $\varrho:\mathbb{R}^{d\times n}$ $\rightarrow$  $\mathbb{T}_{M^{+}}$ that satisfies the following properties.
		\begin{description}
			\item[B0] Normalization: $K_{M}\subseteq \varrho(0)$ and $\varrho(0)\cap -intK_{M}=\phi$;
			\item[B1] Cash losses: for any $z\in K_{M}$, $z\in \varrho(-z1_{n})$;
			\item[B2] Monotonicity: for any $X$,$Y\in \mathbb{R}^{d\times n}$, $X-Y\in K1_{n}$ implies $\varrho(X)\supseteq \varrho(Y)$;
			\item[B3] Loss-dependence: for any $X\in \mathbb{R}^{d\times n}$, $\varrho(X)=\varrho(X\wedge_{K1_{n}}0)$, where
			\[
			X\wedge_{K1_{n}}0:=\left\{ \begin{array}{ll}
			X, & \textrm{$X\notin K1_{n}$},\\
			0, & \textrm{$X\in K1_{n}$}.
			\end{array} \right.
			\]
			\item[B4] Convexity: for any $\lambda\in[0,1]$ and $X$,$Y\in \mathbb{R}^{d\times n}$, $\varrho(\lambda X+(1-\lambda)Y)\supseteq \lambda\varrho(X)+(1-\lambda)\varrho(Y)$.\\
		\end{description}
\end{Definition}

\begin{Remark}
The set-valued loss-based risk statistics start from the point of regulators. Namely, the regulators almost only focus on the loss of investment rather than revenue. Especially, the axiom of translation invariance in coherent and convex risk statistics will definitely fail when we only deal with the loss-based risk. Therefore, the loss-based risk is particularly interesting.
\end{Remark}

 We claim that the set-valued loss-based risk statistics are the special cases of set-valued cash sub-additive risk statistics.
Indeed, for any $X\in \mathbb{R}^{d\times n}$, $z\in K_{M}$ and $\varepsilon\in (0,1)$, we have
\begin{eqnarray*}
	\varrho\Big((1-\varepsilon)X-z1_{n}\Big)&=&\varrho\Big((1-\varepsilon)X+\varepsilon(-\frac{z}{\varepsilon})1_{n}\Big)\\
	&\supseteq&(1-\varepsilon)\varrho(X)+\varepsilon\varrho(-\frac{z}{\varepsilon}1_{n})\\
	&\supseteq&(1-\varepsilon)\varrho(X)+z,
\end{eqnarray*}
where the last inclusion is due to the property of cash losses.
Therefore, by the arbitrariness of $\varepsilon$, we conclude that
\begin{displaymath}
\varrho(X-z1_{n})\supseteq\varrho(X)+z.
\end{displaymath}
This indicates that $\varrho$ satisfies the property $\mathbf{A4}$. Therefore, $\varrho$ is cash sub-additive.\\

Next, we introduce an example of set-valued loss-based risk statistic called $AV@R^{loss}$. Note that, Hamel et al. (2013) first introduced set-valued AV@R, where they also provided the representation result and proved it is a set-valued coherent risk measure. 

\begin{Example}
		(Loss-based average value at risk)
		For any $X\in \mathbb{R}^{d\times n}$ and $0<\alpha<1$, we define $AV@R^{loss}_{\alpha}$ as
		\[
		AV@R^{loss}_{\alpha}(X)
		=\inf_{z\in \mathbb{R}^{d}}\Big\{\frac{1}{\alpha}(-(X\wedge_{K1_{n}}0)|_{M}+z)^{+}-z\Big\}+\mathbb{R}^{m}_{+}.
		\]
\end{Example}

It is clear that $AV@R^{loss}$ satisfies all the properties of Definition~\ref{D2}. So $AV@R^{loss}$ is a set-valued loss-based risk statistic, and hence it is also cash sub-additive. In fact, the AV@R has been widely used in many fields, including bank brokers. And the $AV@R^{loss}$ is just one of his practical applications of set-valued cash sub-additive risk statistics. \\

We now need to derive the representation results of cash sub-additive risk statistics. Any pair $(X, \mu)$, where $X\in \mathbb{R}^{d\times n}$ and $\mu\in M$, can be viewed as the coordinates of a  $\overline{X}\in \mathbb{R}^{d\times n}$ in $\mathcal{T}:=\{0,1\}$ with the element $\theta$,
\begin{equation}\label{32}
\overline{X}(\theta):=XI_{\{1\}}(\theta)+\mu 1_{n}I_{\{0\}}(\theta).
\end{equation}

 For any $\overline{X}$, $\overline{Y}\in \mathcal{X}$ with $\overline{X}=XI_{\{1\}}+\mu_{1}1_{n}I_{\{0\}}$ and $\overline{Y}=YI_{\{1\}}+\mu_{2}1_{n}I_{\{0\}}$, where $X, Y\in \mathbb{R}^{d\times n}$, $\mu_{1},\mu_{2}\in M$, we define the order as $\overline{X}-\overline{Y}\in K1_{n}$ in the case of $Y\leq_{K1_{n}}X$ and $ \mu_{2}\leq_{K} \mu_{1}$.\\

We begin with stating the relation between a set-valued convex risk statistic and a set-valued cash sub-additive risk statistic.

\begin{Proposition}\label{P1}
		Given a set-valued cash sub-additive risk statistic $R$ in $\mathbb{R}^{d\times n}$ with $0\in R(0)$, we define a set-valued risk statistic $\rho$ as follows. For any $\overline{X}$ defined as (\ref{32}), with $X\in \mathbb{R}^{d\times n}$, $\mu \in M$,
		\begin{equation}\label{33}
		\rho(\overline{X}):=R(X-\mu 1_{n})-\mu.
		\end{equation}
		Then $\rho$ is a set-valued convex risk statistic with $\rho(0)=0$ and $\rho(XI_{\{1\}})=R(X)$.
\end{Proposition}

Before we state the representation results of cash sub-additive risk statistics, the representation results of set-valued convex risk statistics should be recalled. The set-valued convex risk statistics were studied by Chen and Hu (2017). We now only state their main results and omit their proofs.

\begin{Proposition}\label{P12}
		$\rho:\mathbb{R}^{d\times n}$ $\rightarrow$ $\mathbb{T}_{M}$ is a set-valued proper closed convex risk statistic in the case of there is a function
		$-\alpha:(K1_{n})^{+}\cap \mathbb{R}^{d\times n}_{+}$ $\rightarrow$ $\mathbb{T}_{M}$ 
		such that for any $\overline{X}\in \mathbb{R}^{d\times n}$,
		\begin{equation}\label{343}
		\rho(\overline{X})=\bigcap_{v\in (K1_{n})^{+}\cap \mathbb{R}^{d\times n}_{+}}\Big\{-\alpha(v)+S_{v}(-\overline{X})\Big\},
		\end{equation}
		where
		\begin{displaymath}
		S_{v}(-\overline{X}):=\{u\in M: v^{tr}(\overline{X}+u1_{n})\geq 0\}.
		\end{displaymath}
		In particular, (\ref{343})is satisfied with $-\alpha$ replaced by $-\alpha_{\min}$ with  
		\begin{displaymath}
		-\alpha_{\min}(v):=cl\bigcup_{\overline{Z}\in \{
			\overline{X}\in \mathbb{R}^{d\times n}: 0\in \rho(\overline{X}) \}}S_{v}(\overline{Z}).
		\end{displaymath}
\end{Proposition}

Using Propositions~\ref{P12}, we are able to derive the representation results of set-valued cash sub-additive risk statistics.\\

\begin{Theorem}\label{T1}
		Any set-valued proper closed cash sub-additive risk statistic $R$ in $\mathbb{R}^{d\times n}$ has the following form. For any $X\in \mathbb{R}^{d\times n}$,
		\begin{equation}\label{35}
		R(X)=\bigcap_{v\in (K1_{n})^{+}\cap \mathbb{R}^{d\times n}_{+}}\Big\{-\gamma(v)+\mathfrak{S}_{v}(-X)\Big\},
		\end{equation}
		where 
		\[
		-\gamma:(K1_{n})^{+}\cap \mathbb{R}^{d\times n}_{+} \rightarrow \mathbb{T}_{M}
		\]
		and
		\begin{displaymath}
			\mathfrak{S}_{v}(-X):=\{u\in M: v^{tr}(XI_{\{1\}}+u1_{n})\geq 0\}.
		\end{displaymath}
		In particular, (\ref{35}) is satisfied with $-\gamma$ replaced by $-\gamma_{\min}$ with  
		\begin{displaymath}
		-\gamma_{\min}(v):=cl\bigcup_{Z\in \{
			X\in \mathbb{R}^{d\times n}: 0\in R(X) \}}\mathfrak{S}_{v}(Z).
		\end{displaymath}
\end{Theorem}

\section{Alternative data-based versions of set-valued cash sub-additive risk measures}
\label{sec:4}
In this section, we develop another framework of data-based versions of cash sub-additive risk measures. This framework is a little different from the previous one. However, almost all the arguments are the same as those in the previous section. Therefore, we only state the corresponding notations and results, and omit all the proofs and relevant explanations.\\

We replace $M $ by $\widetilde{M}\in \mathbb{R}^{d\times n}$ that is a linear subspace of $\mathbb{R}^{d\times n}$. We also replace  $K $ by $\widetilde{K}\in \mathbb{R}^{d\times n}$ that is a closed convex polyhedral cone where $\widetilde{K}\supseteq \mathbb{R}^{d\times n}_{+}$.
The partial order respect to $\widetilde{K}$ is defined as $X\leq_{\widetilde{K}} Y$, which means $Y-X\in \widetilde{K}$.  Denote $\widetilde{K}_{\widetilde{M}}:=\widetilde{K}\cap \widetilde{M}$ by the closed convex polyhedral cone in $\widetilde{M}$, $int\widetilde{K}_{\widetilde{M}}$ the interior of $\widetilde{K}_{\widetilde{M}}$ in $\widetilde{M}$. We denote $\mathbb{T}_{\widetilde{M}}:=\{\widetilde{A}\subset \widetilde{M}:\widetilde{A}=clco(\widetilde{A}+\widetilde{K}_{\widetilde{M}})\}$.\\

We begin with recalling the axioms related to data-based versions of set-valued convex risk measures in Chen and Hu (2017).\\

\begin{Definition}
	A set-valued data-based convex risk statistic is a function $\widetilde{\rho}:\mathbb{R}^{d\times n}$ $\rightarrow$ $\mathbb{T}_{\widetilde{M}}$ that satisfies the following properties,
	\begin{description}
		\item[C0] Normalization: $\widetilde{K}_{\widetilde{M}}\subseteq \widetilde{\rho}(0)$ and $\widetilde{\rho}(0)\cap -int\widetilde{K}_{\widetilde{M}}=\phi$;
		\item[C1] Monotonicity: for any $X_{1}$,$X_{2}\in \mathbb{R}^{d\times n}$, $X_{1}-X_{2}\in \mathbb{R}^{d\times n}\cap \widetilde{K}$ implies that $\widetilde{\rho}(X_{1})\supseteq \widetilde{\rho}(X_{2})$;
		\item[C2] Translative invariance: for any $X\in \mathbb{R}^{d\times n}$ and $z\in \widetilde{M}$, $\widetilde{\rho}(X-z)=\widetilde{\rho}(X)+z$;
		\item[C3]Convexity: for any $X,Y\in \mathbb{R}^{d\times n}$, $\lambda\in[0,1]$, $\widetilde{\rho}(\lambda(X)+(1-\lambda)Y)\supseteq \lambda\widetilde{\rho}(X)+(1-\lambda)\widetilde{\rho}(Y)$.
	\end{description}
\end{Definition}

We now define the set-valued data-based cash sub-additive risk statistics.

\begin{Definition}\label{D41}
		A set-valued data-based cash sub-additive risk statistic is a function $\widetilde{R}:\mathbb{R}^{d\times n}$ $\rightarrow$ $\mathbb{T}_{\widetilde{M}}$ that satisfies $\mathbf{C0},\mathbf{C1},\mathbf{C3}$ and the following property.
		\begin{description}
				\item[C4] Cash sub-additivity: for any $X\in \mathbb{R}^{d\times n}$ and $z\in \widetilde{K}_{\widetilde{M}}$,
				\begin{displaymath}
			\widetilde{R}(X+z)\subseteq \widetilde{R}(X)-z\qquad\textrm{or}\qquad \widetilde{R}(X-z)\supseteq \widetilde{R}(X)+z.
				\end{displaymath}
		\end{description}
\end{Definition}

We need more notions.
Any pair $(X, \widetilde{\mu})$, where $X\in \mathbb{R}^{d\times n}$ and $\widetilde{\mu}\in \widetilde{M}$, can be viewed as the coordinates of a  $\widetilde{X}\in \mathbb{R}^{d\times n}$ in $\mathcal{T}:=\{0,1\}$ with the element $\theta$,
\begin{equation}\label{322}
\widetilde{X}(\theta):=XI_{\{1\}}(\theta)+\widetilde{\mu}I_{\{0\}}(\theta).
\end{equation}

\begin{Proposition}
		Given a set-valued data-based cash sub-additive risk statistic $\widetilde{R}$ in $\mathbb{R}^{d\times n}$ with $0\in \widetilde{R}(0)$, we define a set-valued data-based risk statistic $\widetilde{\rho}$ as follows. For any $\widetilde{X}$ defined as (\ref{322}) where $X\in \mathbb{R}^{d\times n}$, $\widetilde{\mu}\in \widetilde{M}$,
		\begin{equation}
		\widetilde{\rho}(\widetilde{X}):=\widetilde{R}(X-\widetilde{\mu})-\widetilde{\mu}.
		\end{equation}
		Then $\widetilde{\rho}$ is a set-valued data-based convex risk statistic with $\widetilde{\rho}(0)=0$ and $\widetilde{\rho}(XI_{\{1\}})=\widetilde{R}(X)$.
\end{Proposition}

The representation results of set-valued data-based convex risk statistics were studied in Chen and Hu (2017).
We now state the main results of this section.

\begin{Theorem}
		Any set-valued data-based proper closed cash sub-additive risk statistic $\widetilde{R}$ in $\mathbb{R}^{d\times n}$ has the following form. For any $X\in \mathbb{R}^{d\times n}$,
		\begin{equation}\label{411}
		\widetilde{R}(X)=\bigcap_{\widetilde{v}\in \widetilde{K}^{+}\cap \mathbb{R}^{d\times n}_{+}}\Big\{-\widetilde{\gamma}(\widetilde{v})+\widetilde{\mathfrak{S}}_{\widetilde{v}}(-X)\Big\},
		\end{equation}
		where 
		\[
		-\widetilde{\gamma}:\widetilde{K}^{+}\cap \mathbb{R}^{d\times n}_{+} \rightarrow \mathbb{T}_{\widetilde{M}}
		\]
		and
		\begin{displaymath}
		\widetilde{\mathfrak{S}}_{\widetilde{v}}(-X):=\{\widetilde{u}\in \widetilde{M}: \widetilde{v}^{tr}(XI_{\{1\}}+\widetilde{u})\geq 0\}.
		\end{displaymath}
		In particular, (\ref{411})is satisfied with $-\widetilde{\gamma}$ replaced by $-\widetilde{\gamma}_{\min}$ with  
		\begin{displaymath}
		-\widetilde{\gamma}_{\min}(\widetilde{v}):=cl\bigcup_{Z\in \{
			X\in \mathbb{R}^{d\times n}: 0\in \widetilde{R}(X) \}}\mathfrak{S}_{\widetilde{v}}(Z).
		\end{displaymath}
\end{Theorem}

\section{Proofs of main results}
\label{sec:5}
In this section, we provide all the proofs of the results stated in Sect.~\ref{sec:3}.\\

 \noindent \textbf{Proof of Proposition~\ref{P1}. }
 It is easy to check that $\rho(0)=0$, $\rho(XI_{\{1\}})=R(X)$ and $\rho$  satisfies the property of $\mathbf{A0}$. Next, we derive that $\rho$ satisfies properties of $\mathbf{A1},\mathbf{A2}$ and $\mathbf{A3}$.
 \begin{description}
 	\item[A1.] Monotonicity: for any 
 	$\overline{X}=XI_{\{1\}}+\mu_{1}1_{n}I_{\{0\}}$ and $\overline{Y}=YI_{\{1\}}+\mu_{2}1_{n}I_{\{0\}}$, where $X, Y\in \mathbb{R}^{d\times n}$, $\mu_{1},\mu_{2}\in M$ with
 	$\overline{X}-\overline{Y}\in K1_{n}$, then
 	\begin{displaymath}
 	\rho(\overline{X})=R(X-\mu_{1}1_{n})-\mu_{1}\supseteq R(X-\mu_{2}1_{n})-\mu_{2}\supseteq R(Y-\mu_{2}1_{n})-\mu_{2}=\overline {\rho}(\overline{Y}),
 	\end{displaymath}
 	which shows that $\rho$ is monotone.
 	\item[A2.] Translative invariance: for any $b\in M$,  $\overline{X}=XI_{\{1\}}+\mu1_{n}I_{\{0\}}$ where $X\in \mathbb{R}^{d\times n}$ and $\mu\in M$,
 	\begin{eqnarray*}
 		\rho(\overline{X}+b1_{n})&=&\rho\Big((X+b1_{n})I_{\{1\}}+(\mu 1_{n}+b1_{n})I_{\{0\}}\Big)\\
 		&=&R\Big(X+b1_{n}-(\mu+b)1_{n}\Big)-\mu-b\\&=&R(X-\mu 1_{n})-\mu-b\\&=&\rho(\overline{X})-b,
 	\end{eqnarray*}
 	which shows that $\rho$ satisfies the translative invariance.
 	\item[A3.] Convexity: for any  $\lambda\in (0,1)$, $\overline{X}=XI_{\{1\}}+\mu_{1}1_{n}I_{\{0\}}$ and $\overline{Y}=YI_{\{1\}}+\mu_{2}1_{n}I_{\{0\}}$, where $X, Y\in \mathbb{R}^{d\times n}$, $\mu_{1},\mu_{2}\in M$,
 	\begin{eqnarray*}
 		&&\rho(\lambda\overline{X}+(1-\lambda)\overline{Y})\\&=& \rho\Big(\big(\lambda{X}+(1-\lambda){Y}\big)I_{\{1\}}+\big(\lambda{\mu_{1}}+(1-\lambda){\mu_{2}}\big)1_{n}I_{\{0\}}\Big)\\
 		&=&{R}\Big(\big(\lambda{X}+(1-\lambda){Y}\big)-\big(\lambda{\mu_{1}}+(1-\lambda){\mu_{2}}\big)1_{n}\Big)-\lambda{\mu_{1}} -(1-\lambda){\mu_{2}}\\
 		&=&{R}\Big(\lambda(X-\mu_{1}1_{n})+(1-\lambda)(Y-\mu_{2}1_{n})\Big)-\lambda{\mu_{1}}-(1-\lambda){\mu_{2}}\\
 		&\supseteq&\lambda R(X-\mu_{1}1_{n})+(1-\lambda)R(Y-\mu_{2}1_{n})-\lambda{\mu_{1}}-(1-\lambda){\mu_{2}}\\
 		&=&\lambda\rho(\overline{X})+(1-\lambda)\rho(\overline{Y}),
 	\end{eqnarray*}
 	which shows that $\rho$ is convex.
 \end{description}
\qed

\noindent \textbf{Proof of Proposition~\ref{P12}. }
See Chen and Hu (2017). \\
\qed

\noindent \textbf{Proof of Theorem~\ref{T1}. }
From Proposition~\ref{P1}, we can define a set-valued convex risk statistic $\rho$ in $\mathbb{R}^{d\times n}$ by $R$, such that
$\rho(XI_{\{1\}})=R(X)$ for any $X\in \mathbb{R}^{d\times n}$. Indeed, as $0\in \mathbb{R}^{d}$, we have $XI_{\{1\}}\in \mathbb{R}^{d\times n}$. Therefore, from Proposition~\ref{P12},
\begin{displaymath}
R(X)=	\rho(XI_{\{1\}})=\bigcap_{v\in (K1_{n})^{+}\cap \mathbb{R}^{d\times n}_{+}}\Big\{-\alpha(v)+S_{v}(-XI_{\{1\}})\Big\},
\end{displaymath}
where $-\alpha:(K1_{n})^{+}\cap \mathbb{R}^{d\times n}_{+}$ $\rightarrow$ $\mathbb{T}_{M}$ and
\begin{displaymath}
S_{v}(-\overline{X}):=\{u\in M: v^{tr}(\overline{X}+u1_{n})\geq 0\}.
\end{displaymath}
In particular, the function $-\alpha$ can be replaced by $-\alpha_{\min}$ where  
\begin{displaymath}
-\alpha_{\min}(v):=cl\bigcup_{Z\in \{
	X\in \mathbb{R}^{d\times n}: 0\in \rho(XI_{\{1\}}) \}}S_{v}(ZI_{\{1\}}).
\end{displaymath}
We now denote 
\[
\mathfrak{S}_{v}(X):=\{u\in M: v^{tr}(-XI_{\{1\}}+u1_{n})\geq 0\}.
\]
Therefore, the set-valued cash sub-additive risk statistic $R$ can be expressed as
\[
R(X)=\bigcap_{v\in (K1_{n})^{+}\cap \mathbb{R}^{d\times n}_{+}}\Big\{-\gamma(v)+\mathfrak{S}_{v}(-X)\Big\},
\]
where $-\gamma:(K1_{n})^{+}\cap \mathbb{R}^{d\times n}_{+} \rightarrow \mathbb{T}_{M}$ and can replaced by $-\gamma_{\min}$ where
\[
-\gamma_{\min}(v):=cl\bigcup_{Z\in \{
	X\in \mathbb{R}^{d\times n}: 0\in R(X) \}}\mathfrak{S}_{v}(Z).
\] 
\qed

\section*{Conclusions} 
In fact, the time value of money is a critical factor in insurance and financial applications. However, in traditional research of risk statistic, the time value of money is always ignored, that is, when $m$ dollars are added to a future position, the capital requirement at time $t=0$ is still $m$ dollars. This is obviously inconsistent with reality.
Thus, we derive a new class of risk statistics, named set-valued cash sub-additive risk statistic. 
Yet, we do not conduct theoretical analysis on exponential dispersion models like Shushi and Yao (2020).
Our results provide the macro models from the perspective of the time value of money.


%
%



\end{document}